\documentstyle[preprint,epsfig,aps]{revtex}
\begin{document}

\title{Neutrino mediated muon--electron conversion in nuclei revisited}
\author{ 
F. \v Simkovic
\footnote{On  leave of absence from 
Department of Nuclear Physics,  Comenius University, 
Mlynsk\'a dolina F1, SK--842 15 Bratislava, Slovakia}, 
V.E. Lyubovitskij, Th. Gutsche, Amand Faessler
}
\address{
Institut f\"ur Theoretische Physik, Universit\"at 
T\"ubingen, Auf der Morgenstelle 14, D-72076 T\"ubingen, Germany}
\author{ Sergey Kovalenko
\footnote{On leave of absence from 
the Joint Institute for Nuclear Research, Dubna, Russia}}
\address{ Departamento de F\'\i sica, Universidad T\'ecnica Federico 
Santa Mar\'\i a, Casilla 110-V, Valpara\'\i so, Chile} 
%

\maketitle 

\draft 

\begin{abstract}
The non-photonic neutrino exchange mechanism of the lepton flavor violating
muon--electron conversion  in nuclei is revisited.
First we determine the nucleon coupling constants for
the neutrino exchange mechanism in a relativistic quark model
taking into account quark confinement and chiral symmetry requirements.
This includes a new, previously overlooked tree-level contribution from  
neutrino exchange between two quarks in the same nucleon.  
Then for the case of an additional sterile  neutrino we reconsider 
the coherent mode of this process.  
The presence of a mixed sterile-active neutrino state 
$\nu_h$ heavier than the quark confinement scale $\Lambda_c \sim 1~{\rm GeV}$  
may significantly improve the prospects for observation of this process in 
future experiments as compared to  the conventional scenario 
with only light neutrinos.  
Turning the arguments around we derive new experimental constraints on  
$\nu_h-\nu_{e,\mu}$ mixing from the non-observation of 
muon--electron conversion.  
\end{abstract}

The recent results from the Super-Kamiokande \cite{SK-global} 
and SNO \cite{sno} experiments on atmospheric and 
solar neutrinos give convincing evidence on neutrino oscillation
 and hence on neutrino masses and lepton flavor violation (LFV). 
As is known,  experimental searches for rare processes offer complimentary 
information on the LFV. This can shed additional 
light on the physics underlying this phenomenon, discriminating various models
beyond the standard model (SM) (for a review see \cite{ko94}). 
The muon--electron [$(\mu^-,e^-)$] 
conversion in nuclei \cite{ko94,marc,ko97,mue,kos01,sim97}, 
\begin{equation} 
\mu^-_b + (A,Z) \rightarrow (A,Z) + e^-, 
\label{eq.1} 
\end{equation} 
is one of the most prominent lepton flavor changing reaction. Experiments searching 
for this process in the coherent channel with a monoenergetic final state electron have 
reached an unprecedented level of sensitivity.   
Presently, the most stringent  upper limits on the branching ratio 
$R_{\mu e}$ related to the $(\mu^-,e^-)$ conversion 
has been set by the  SINDRUM II collaboration \cite{sind}: 
\begin{eqnarray} 
R_{\mu e} = \frac{\Gamma_{\mu e}}{\Gamma_{\mu \nu_\mu}} 
< 6.1\times 10^{-13} ~~(target:~^{48}Ti), ~~~
2.0\times 10^{-11} ~~(target:~^{79}Au), 
\end{eqnarray} 
where ${\Gamma_{\mu e}}$ and ${\Gamma_{\mu \nu_\mu}}$ are the rates of 
the $(\mu^-,e^-)$ conversion  and ordinary muon capture, respectively. 
Future experiments will significantly improve these limits.
There are proposals of the SINDRUM II collaboration to reduce 
the current limits on the  ratio $R_{\mu e}$
for $^{48}Ti$ and $^{197}Au$  down to 
$10^{-14}$ and $6\times 10^{-13}$ \cite{sind}, respectively. 
A new Muon Electron COnversion (MECO) experiment on $^{27}Al$ 
is  planned at BNL \cite{meco} with an expected  sensitivity 
on the branching ratio of about $2\times10^{-17}$. 
Another future project PRIME \cite{prime} for the $(\mu^-,e^-)$ conversion 
on $^{48}Ti$ is going to reach a sensitivity of $10^{-18}$. 
The realization of these projects would allow to  set 
new stringent constraints on the LFV interactions relevant for 
the $(\mu^-,e^-)$ conversion. 
This process can be triggered by the LFV interactions associated with 
the exchange of neutrinos 
and/or new heavy particles (neutralinos, charginos, leptoquarks etc.) 
predicted in models beyond the SM \cite{ko94,marc,ko97,mue,kos01}. 
In general all the $(\mu^-,e^-)$ conversion mechanisms can be separated 
into photonic and non-photonic ones. 
A mechanism is photonic if it involves a virtual photon line connecting 
the effective leptonic LFV current with the electromagnetic nuclear current, 
otherwise a mechanism is non-photonic. These classes of mechanisms 
differ significantly 
on their particle and nuclear physics sides and are usually studied independently.

In this letter we concentrate on the non-photonic neutrino exchange mechanism.
We are studying a model with  three left-handed, weak doublet neutrinos 
$\nu'_{Li} = (\nu'_{Le},\nu'_{L\mu},\nu'_{L\tau})$ and a certain number $n$ of 
the SM singlet, right-handed sterile neutrinos $\nu'_{Ri}=(\nu'_{R1},...\nu'_{Rn})$. 
Due to mixing they form $n+3$ neutrino mass eigenstates $N_i$ with 
masses $m_{i}$  related  to the weak eigenstates
$\nu'_{\alpha} = (\nu'_{L e}, \nu'_{L \mu}, \nu'_{L \tau}, 
\nu_{R1}^{\prime c},..., \nu_{Rn}^{\prime c})$  
by an unitary mixing matrix $U$ as
$
N_{i} = U^*_{\alpha i} \nu'_{\alpha}.
$

Among $N_i$ there must be at least three observable light neutrinos dominated by 
the active $\nu'_{e,\mu,\tau}$ components while the other states 
may be of arbitrary mass.  In particular, they may include additional light neutrinos
($\nu_i$),
one of which might be relevant for the  
phenomenology of neutrino oscillations , 
as well as intermediate and heavy  mass neutrinos ($\nu_{_h}$). 
The presence 
or absence of these neutrino states is an issue for experimental searches.
For simplicity we consider the  $(\nu_i, \nu_h)$ scenario with an arbitrary number of light 
neutrinos $\nu_i$ with  masses $m_{\nu i}$ on the eV scale
and one neutrino state $\nu_h$ with 
mass $m_h$ larger than the typical hadronic scale $\Lambda_c \sim 1$ GeV. 
A similar scenario, but with $\nu_h$ having a mass below $1$ GeV was previously studied 
in connection with $K^{\pm}$ and $\tau$ semileptonic decays \cite{DGKS}.

\begin{figure}
\vbox{
\centerline{\epsfig{file=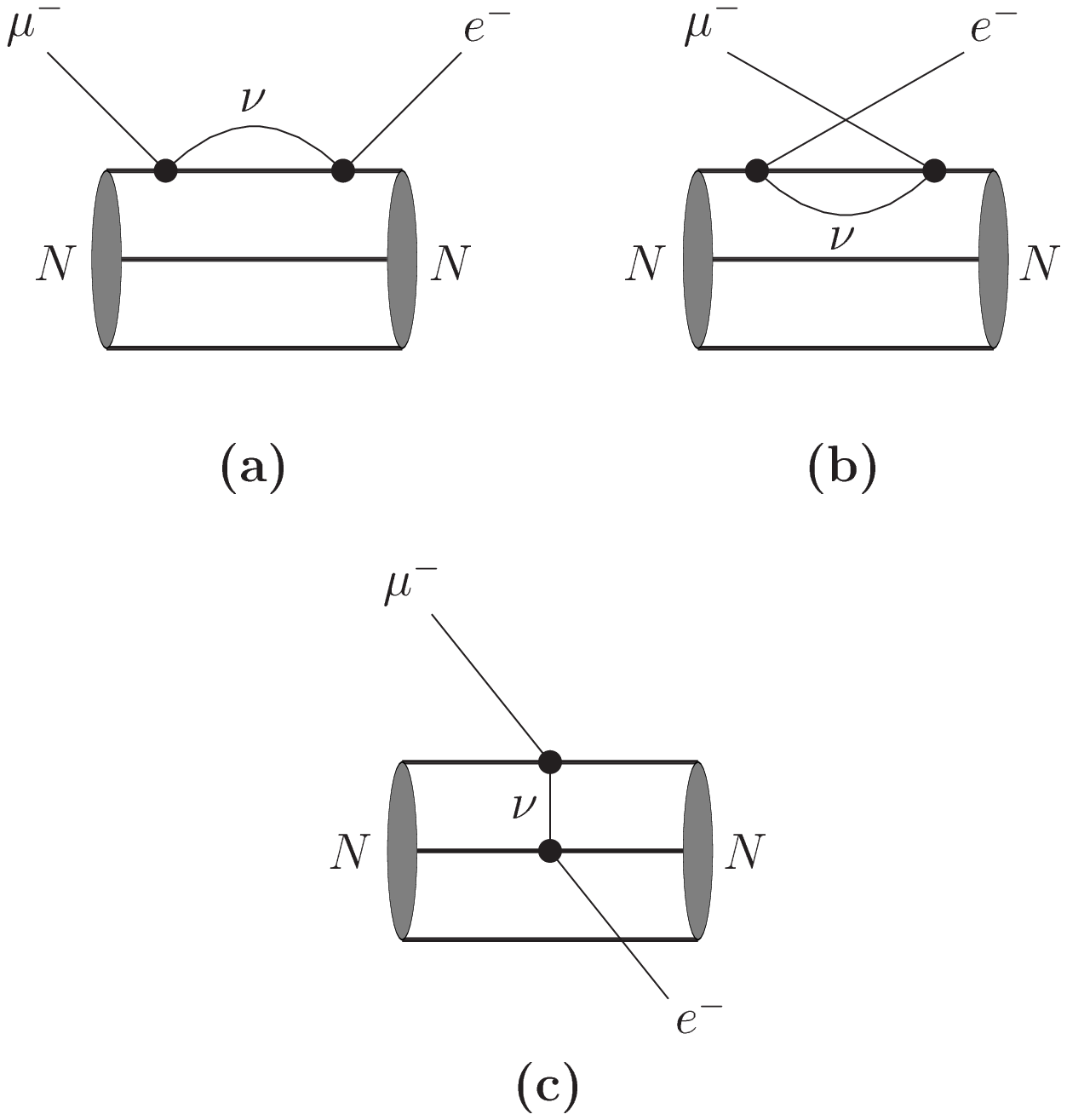,height=26.cm,width=18.cm} }
\vspace{-6.0cm}
}
\caption{Feynman diagrams for $(\mu^-,e^-)$ 
conversion in nuclei associated with  neutrino exchange
on the quark level. 
Both one- (a,b) and two-body (c) mechanisms are considered. }
\label{fig.1}
\end{figure}

The analysis of the nuclear $(\mu^-,e^-)$ conversion starts with the 
elementary nucleon process $\mu^- + N \rightarrow N + e^-$. In models 
with  non-trivial neutrino mixing this process can be realized at the 
quark level according to the diagrams of Fig. \ref{fig.1}. The diagrams of
Fig. 1a,b are the well-known one-quark box diagrams \cite{ko94} 
while that of Fig. 1c is the new tree-level two-quark diagram. These 
lowest order diagrams represent the complete set of the neutrino 
exchange diagrams on quark-level  relevant for the above nucleon process. 

In the process considered
the typical momentum  transfer $Q^2$ to the nucleon 
is small comparable to the scale set by the muon mass with
$Q^2\sim m^2_{\mu}$. Therefore, 
the quarks in the  diagrams considered
cannot be treated as free particles as would be the case 
in the asymptotic region $Q^2 \gg \Lambda^2_c$. An appropriate 
treatment should deal with quarks as states which are 
confined in the  nucleon. Due to 
the lack of a rigorous theory for confinement in QCD one has to engage 
phenomenological models. In this work we 
are using the perturbative chiral quark model (PCQM) 
\cite{Gutsche,PCQM} treating quarks as extended objects, the constituent 
quarks, which are confined in the nucleon. In this model each quark vertex 
acquires a form factor with the characteristic momentum scale  
$\Lambda_c \sim 1$ GeV related to the confinement length $l_c \sim \Lambda_c^{-1}$. 
In the diagrams of Fig. 1 these form factors 
set the scale  for the loop momentum $q_{\nu}$ of the virtual neutrino. 
This is in  contrast to the previous analysis \cite{ko94} of the diagrams  Fig. 1a,b
associated with different (neutrino mass dependent) terms of the 
$(\mu^-,e^-)$ conversion amplitude, 
where the $q_{\nu}$ scale is set by  the W-boson mass due to the presence of the 
corresponding  propagators  in these diagrams.  
 
Knowing the characteristic scale $q_0 \sim \Lambda_c$ of the neutrino momentum $q_{\nu}$ in 
the diagrams of Fig. 1, we consider the general structure of
the $(\mu^-,e^-)$ conversion amplitude ${\cal A}_{\mu e}$. 
In the  $(\nu_i, \nu_h)$ neutrino scenario introduced above
one can write
\begin{eqnarray}    
{\cal A}_{\mu e} &\sim&
\int \left(\sum_i~ \frac{U_{\mu i} U^*_{e i}}{q^2 - m_i^2 + i \epsilon}\right) 
\cdot G(q^2/q_0^2) d^4 q\\ \nonumber   
&\sim& 
\int \frac{1}{q^2}~ \sum^{light}_i~ {U_{\mu i} U^*_{e i}}~ 
(1~ +~ \frac{m^2_i}{q^2}~ ... ~)\cdot G(q^2/q_0^2) d^4 q -
\frac{U_{\mu h} U^*_{e h}}{m_h^2} \int G(q^2/q_0^2) d^4 q
~(m_h \gg q_0).
\label{eq.2}
\end{eqnarray}
Here $G(q^2/q^2_0)$ is a characteristic 
function suppressing the contribution for $q^2 \gg q^2_0$ 
in the loop momentum integration. Then it follows that
\begin{eqnarray}
{\cal A}_{\mu e} \approx 
\left\{
\begin{array}{l}
\left(\sum_i~ U_{\mu i} U^*_{e i} \frac{m_i^2}{q_0^2} \right) 
\int \frac{q_0^2}{q^2} G(q^2/q_0^2) \frac{d^4 q}{q^2}, \ \ \  \mbox{for}  \ \ \ m_h \ll q_0\\
- U_{\mu h} U^*_{e h}  \int G(q^2/q_0^2) \frac{d^4 q}{q^2} \ \ \ \ \ \ \ \ \ \ \mbox{for} \ \ \ m_h \gg q_0
\end{array}\right.
\label{eq.3}
\end{eqnarray}
as a consequence of the unitarity of the mixing matrix with
$\sum_i U_{\mu i} U^*_{e i} = 0$.
 
Previously, only the case $\ m_h \ll q_0$ of Eq. (\ref{eq.3}) 
was considered in the literature \cite{ko94,ko97,vergr}. 
Because of the smallness of the  
ratio $m_i^2/q_0^2$ the neutrino exchange mechanism 
leads to  rates for the $(\mu^-,e^-)$ conversion which are out of 
reach for  ongoing and near 
future experiments. The situation changes if there exists a 
heavy neutrino state $\nu_h$ 
with mass $\ m_h \gg q_0$ and with a non-vanishing admixture of active flavors 
$\nu_{\mu, e}$.  In this case the suppression 
factor associated with the small neutrino masses is replaced by 
the product of  mixing matrix 
elements $U_{\mu h} U^*_{e h}$ as indicated
in  Eq. (\ref{eq.3}). 
Since the existing experimental constraints on $U_{\mu h} U^*_{e h}$ are not 
stringent \cite{PDG} one may expect much larger rates for 
$(\mu^-,e^-)$ conversion 
in the $(\nu_l, \nu_h)$ scenario than for the case without 
an intermediate mass $\nu_h$ state. 

Following the standard 
approach \cite{kos01} we consider the effective nucleon Lagrangian written as 
\begin{eqnarray} 
{\cal L}^{eff.}_{\mu e}(x)~ &=&~ 
G^2_F~ m^2_\mu ~  U_{\mu h} U^*_{e h} ~ \overline{e}(x) 
\gamma_\alpha (1-\gamma_5) \mu(x) \times \nonumber \\ 
&& [ \overline{p}(x) \gamma^\alpha (f^p_V - f^p_A \gamma_5) p(x) + 
  \overline{n}(x) \gamma^\alpha (f^n_V - f^n_A \gamma_5) n(x) ] + h.c.. 
\label{eq.4} 
\end{eqnarray}
Here $m_\mu$ is the mass of the muon.
The  partial contributions of the diagrams  Fig. 1a,b,c
are contained in the coupling coefficients as
$f_{V,A}^{N} = f_{V,A}^{N;1a} + f_{V,A}^{N;1b} + f_{V,A}^{N;1c}$. 
In the present work we restrict ourselves to the dominant mode of  $(\mu^-,e^-)$ conversion, 
where the axial-vector current contribution \cite{vergr} is neglected and, thus, 
only the vector form factors $f_{V}^{p,n}$ are relevant for the subsequent analysis.

We evaluated the form factors $f_{V}^{p,n}$ within the perturbative chiral quark 
model (PCQM), a relativistic quark model suggested in \cite{Gutsche} and 
extended in \cite{PCQM} for the study of low-energy properties of 
baryons. The model operates with relativistic quark wave functions and 
takes into account  quark confinement as well as chiral symmetry 
requirements. The PCQM was successfully applied to $\sigma$-term physics 
and to the electromagnetic properties of the nucleon \cite{PCQM}. 
In the present analysis we included the contributions from both the one-body 
(Fig. \ref{fig.1}a, b) and the two-body (Fig. \ref{fig.1}c) diagrams neglecting 
the external three-momenta of the  leptons. For the one-body diagrams of 
Fig. \ref{fig.1}a and Fig. \ref{fig.1}b we restrict the expansion of the quark propagator 
to the ground state eigenmode:
\begin{eqnarray}\label{quark_propagator_ground}  
iG_\psi(x,y) \to iG_0(x,y) = u_0(\vec{x}) \bar u_0(\vec{y})  
e^{-i{\cal E}_0(x_0-y_0)}\theta(x_0-y_0), 
\end{eqnarray}
where  ${\cal E}_0$ and $u_0(\vec{x})$ are the quark ground state energy 
and wave function; that is we restrict the intermediate baryon 
states to $N$ and $\Delta$ configurations. In Ref. \cite{PCQM} we showed that 
this approximation for the quark propagator works quite well in the 
phenomenology of low-energy nucleon physics.  

With above approximations 
the partial contributions of the diagrams of Fig. 1a,b,c to the 
coupling constant of the vector current are: 
\begin{eqnarray}\label{Fig1a}
f_V^{N; 1a} &=& \frac{1}{2} \int \frac{d^3k}{(2\pi)^3} \int d^3x \int d^3y\,\, 
e^{i \vec{k}(\vec{x}-\vec{y})} 
\frac{1}{|\vec{k}| - m_\mu - i\varepsilon}\\ \nonumber
&\times&<N|\sum\limits_{i=1}^3 \biggl(\Phi_0(\vec{x}) \Phi_0(\vec{x}) + 
\vec{\Phi}(\vec{x}) \vec{\Phi}(\vec{x}) \biggr)^{(i)} |N>, \\
\label{Fig1b}
f_V^{N; 1b} &=& \frac{1}{2} \int\frac{d^3k}{(2\pi)^3} \int d^3x \int d^3y \,\, 
e^{i \vec{k}(\vec{x}-\vec{y})} \frac{1}{|\vec{k}| + m_\mu} \\  \nonumber
&\times&<N|\sum\limits_{i=1}^3 \biggl(\Phi_0(\vec{x}) \Phi_0(\vec{x}) + 
\vec{\Phi}(\vec{x}) \vec{\Phi}(\vec{x}) \biggr)^{(i)} |N>,  \\
\label{Fig1c}
f_V^{N; 1c} &=& \frac{1}{2} \int \frac{d^3k}{(2\pi)^3} 
\int d^3x \int d^3y \,\, e^{i \vec{k}(\vec{x}-\vec{y})} 
\biggl(\frac{1}{|\vec{k}| - m_\mu - i\varepsilon} -  
\frac{1}{|\vec{k}| + m_\mu} \biggr)\\ \nonumber
&\times&<N|\sum\limits_{i\,\not\! = j}^3 \biggl(\Phi_0(\vec{x})^{(i)} 
\Phi_0(\vec{x})^{(j)} + 
\vec{\Phi}(\vec{x})^{(i)}  \vec{\Phi}(\vec{x})^{(j)}\biggr) |N>. 
\end{eqnarray}
Here $i$ and $j$ are the quark indices, 
$\Phi_0(\vec{x}) = \bar u_0(\vec{x}) \gamma_0 u_0(\vec{x})$ and 
$\vec{\Phi}(\vec{x}) = \bar u_0(\vec{x}) \vec{\gamma} u_0(\vec{x})$ 
are the time and spatial components of the quark vector current. 
 The single components are projected onto the 
three-quark state building up the nucleon state $|N>$.
As in Refs. \cite{PCQM} we use 
the variational {\it Gaussian ansatz} \cite{Duck} for the quark ground 
state wave function given by: 
\begin{eqnarray}\label{Gaussian_Ansatz} 
u_0(\vec{x}) \, = \, N \, \exp\biggl[-\frac{\vec{x}^{\, 2}}{2R^2}\biggr] \, 
\left(
\begin{array}{c}
1\\
i \rho \, \frac{\displaystyle{\vec{\sigma}\vec{x}}}{\displaystyle{R}}\\
\end{array} 
\right) \, \chi_s \, \chi_f\, \chi_c \, , 
\end{eqnarray}      
where $\chi_s$, $\chi_f$, $\chi_c$ refer to the spin, flavor and color 
spinors. The constant $N=[\pi^{3/2} R^3 (1+3\rho^2/2)]^{-1/2}$ is fixed by 
the normalization condition. 

Our Gaussian ansatz contains two model parameters: 
the dimensional parameter $R$ and the dimensionless parameter $\rho$. 
The parameter $\rho$ can be related to the axial coupling constant $g_A$ 
calculated in zeroth-order (or 3q-core) approximation: 
\begin{eqnarray}\label{g_A}
g_A=\frac{5}{3} \biggl(1 - \frac{2\rho^2} {1+\frac{3}{2} \rho^2} \biggr) = 
\frac{5}{3} \frac{1+2\gamma}{3} ,  
\end{eqnarray}
where $\gamma$ is a relativistic reduction factor  
\begin{eqnarray}
\gamma=\frac{1-\frac{3}{2}\rho^2}{1+\frac{3}{2}\rho^2}=
\frac{9}{10}g_A-\frac{1}{2} .
\end{eqnarray}
The parameter $R$ can be physically understood as the mean radius of the 
three-quark core and is related to the charge radius $<r^2_E>^P_{LO}$ 
of the proton in the leading-order (or zeroth-order) approximation 
as \cite{PCQM} 
\begin{eqnarray}\label{r2ep_LO}
<r^2_E>^P_{LO} \, = \, \frac{3R^2}{2} \, 
\frac{1 \, + \, \frac{5}{2} \, \rho^2}
{1 \, + \, \frac{3}{2} \, \rho^2} \, = \, 
R^2 \biggl( 2 - \frac{\gamma}{2} \biggr) . 
\end{eqnarray}
In our calculations we use the tree-level value $g_A$=1.25 as obtained in  
Chiral Perturbation Theory 
\cite{Gasser1} and the averaged value of $R=0.6$ fm \cite{PCQM}
corresponding to $<r^2_E>^P_{LO}=0.6$ fm$^2$. 

A straightforward analytical evaluation of the expressions 
in Eqs. (\ref{Fig1a})-(\ref{Fig1c}) results in 
the following values for the partial isospin dependent
vector coupling constants: 
\begin{eqnarray}
& &f_V^{p; 1a} = 2.37 + i \, 0.41 , ~~~~~f_V^{n; 1a} =  1.19 + i \, 0.21 ,\\
& &f_V^{p; 1b} = 0.64          ,    
~~~~~~~~~~~~~~~~f_V^{n; 1b} =  1.27 , \nonumber\\
& &f_V^{p; 1c} = 0.44 + i \, 0.14 , ~~~~~f_V^{n; 1c} =  0.44 + i \, 0.14 , 
\nonumber 
\end{eqnarray}
For the total 
nucleon vector coupling constants $f_{V}^{p,n}$, entering 
in Eq. (\ref{eq.4}), we finally obtain   the values
\begin{equation}
f^p_V = 3.45 + i \, 0.55, ~~~~~f^n_V = 2.90 + i \, 0.35, 
\label{eq.7}
\end{equation}
which are just the sum of the partial contributions.

Starting from the effective Lagrangian  of 
Eq. (\ref{eq.4}) the branching ratio 
of the coherent $(\mu^-, e^-)$ conversion is derived as 
\begin{eqnarray} 
R_{\mu e} =  |U_{\mu h} U^*_{e h}|^2 ~\frac{2}{\pi} ~m_\mu ~(G_F m^2_\mu)^4 
~\frac{p_e E_e}{m^2_\mu} 
~F(Z,p_e) ~\frac{|{\cal M}_{\mu e}|^2}{\Gamma_{\mu \nu_\mu}}. 
\label{eq.5}
\end{eqnarray} 
Here, $E_e$ ($E_e = m_\mu -\varepsilon_b$, $\varepsilon_b$ is the muon 
binding energy) and $p_e$ ($p_e = |{\vec{p}}_e| $) are  energy and  
momentum of the outgoing electron. $F(Z,p_e)$ is the 
relativistic Coulomb factor \cite{DOI85} and
 the nuclear structure factor is defined as
\begin{equation}
{\cal M}_{\mu e} = \frac{1}{\sqrt{m^3_\mu}} 
(f_V^p {\cal M}_p + f_V^n {\cal M}_n). 
\label{eq.6}
\end{equation}
In our analysis we used values for the nuclear matrix elements
${\cal M}_{p,n}$ as derived in  
Ref. \cite{kos01}. 
Using the calculated vector coupling constants as an input,
in Table \ref{table.1} we indicate the numerical values for the relevant 
quantities entering in Eq.  (\ref{eq.5}). With these values we get the
following results for the coherent $(\mu^-, e^-)$ conversion  
on the nuclear targets
$^{48}Ti$, $^{197}Au$ and $^{27}Al$ as: 
\begin{equation}
\frac{R^{theor}_{\mu e}}{|U_{\mu h} U^*_{e h}|^2}
= 3.19\times 10^{-11} (^{27}Al), ~~ 
6.34\times 10^{-11} (^{48}Ti), ~~   
4.73\times 10^{-10} (^{197}Au), 
\label{eq.8}
\end{equation}
which is valid for a heavy neutrino state $\nu_h$ with mass 
$m_h \gg \Lambda_c \sim 1$ GeV.
From these results we derive  upper limits on the product of 
mixing matrix 
elements $|U_{\mu h} U^*_{e h}|$ which correspond to the 
sensitivity of present and near future 
experiments discussed in the introduction. 
These limits which are set by the experimental upper bounds 
are listed in Table \ref{table.2}. The present 
constraint $|U_{\mu h} U^*_{e h}| \leq 0.1$, provided 
by the SINDRUM II $^{48}Ti$ experiment, is rather weak. 
An improvement on this bound is expected from the ongoing 
SINDRUM II $^{197}Au$ experiment. 
Significant sensitivity, down to $10^{-4}-10^{-3}$, will be hopefully 
achieved by 
the future MECO (target $^{27}Al$) and PRIME (target $^{48}Ti$)
experiments. 

The mixing of massive neutrinos, like $\nu_h$  with 
active $\nu_{e,\mu,\tau}$ flavors 
was previously looked for in various experiments except 
for $(\mu^-,e^-)$ conversion.
An extensive list for the constraints on the
$|U_{e h}|$ and $|U_{\mu h}|$ mixing matrix 
elements for various masses $m_h$ is given in Ref. \cite{PDG}. 
For $m_h \le 19.6$ GeV all the values of $|U_{\mu h}|$  and $|U_{e h}|$  
have been excluded by the MARK II collaboration. The ALEPH collaboration ruled
out all values of these matrix elements for 
$25.0 \le m_h \le 42.7$ GeV. However, above $42.7$ GeV there exist only 
narrow domains of $m_h$ where $|U_{\alpha h}|$, with $\alpha = e, \mu$, are constrained. 
The typical constraints are $|U_{\alpha h}|^2 \le 10^{-10}-10^{-5}$. 
The $(\mu^-,e^-)$  conversion constraints in Table \ref{table.2} 
cover the partially constrained region  $m_h \ge 42.7$ GeV and extend into the region 
$m_h \ge 1600$ GeV currently unconstrained.  On the other hand, 
the poorly explored region      
$m_h \ge 42.7$ GeV offers a loop-hole for observation of 
the $(\mu^-,e^-)$  conversion 
in future experiments as its rate is  unconstrained  in  the most 
part of this mass region.

In summary, we have studied the non-photonic neutrino exchange mechanism of 
coherent $(\mu^-,e^-)$ conversion in nuclei in the presence of sterile 
neutrinos. We found a new tree level contribution to the $(\mu^-,e^-)$ conversion 
(Fig. 1c) which is as important as the previously known box-type contributions (Fig. 1a,b).    
The nucleon form factors, parameterizing the effective 
nucleon Lagrangian, have been
analyzed within the perturbative chiral quark model \cite{PCQM}. In this model 
the momentum scale of the virtual neutrino is set by the quark 
confinement with 
$\Lambda_c \sim 1$ GeV in  the three diagrams. This significantly differs
from the previous analysis \cite{ko94} of the diagrams Fig. 1a,b where 
this scale is of the order of $\sim M_W$. 
The lowering the neutrino momentum 
scale has a notable effect on the analysis of  the observability of 
the $(\mu^-,e^-)$ conversion.
We have shown that in the neutrino scenario with at least one heavy neutrino 
state $\nu_h$ with mass $m_h \gg \Lambda_c \sim 1$ GeV 
the rate of this lepton flavor violating process could be large enough to be observed 
in planed experiments. 
This observation is in 
contrast to the  conventional belief that  $(\mu^-,e^-)$ conversion 
will not be detected
experimentally  even in the distant future if the process is dominated by 
the neutrino mechanism. It turn we derived new upper 
bounds on the product of $\nu_h$ neutrino mixing matrix elements from 
the non-observation of this process in running and planed experiments. 

{\it Acknowledgements}. This work was supported by the Deutsche
Forschungsgemeinschaft (DFG, grants FA67/25-1 and 436 SLK 17/298.)
and in part by Fondecyt (Chile) under grant
8000017, by a C\'atedra Presidencial (Chile) and by RFBR (Russia) under
grant 00-02-17587."


\begin{table}
\caption{The nuclear structure factor ${\cal M}_{\mu e}$ (dimensionless)  and other 
quantities of Eq. (\protect\ref{eq.5}). The experimental values of 
the total muon capture rate $\Gamma_{\mu \nu_\mu}$ are from Ref. [20].
} 
\label{table.1}
\begin{center}
\begin{tabular}{rccccc}
Nucleus & $p_e \, (fm^{-1})$ & $\epsilon_b \, (MeV)$ &
$\Gamma_{\mu \nu_\mu} \, ( \times 10^{6} \, s^{-1})$ &
$F(Z,p_e)$ & $|{\cal M}_{\mu e}|$  \\
\hline
$^{27}$Al  & 0.531 &  -0.470 &  0.71 & 1.37 & 0.754   \\
$^{48}$Ti  & 0.529 &  -1.264 &  2.60 & 1.62 & 1.87   \\
$^{197}$Au & 0.485&   -9.938 & 13.07 & 3.91 & 7.37  \\
\end{tabular}
\end{center}
\end{table}

\begin{table}[t]
\caption{Upper bounds on the product of mixing matrix elements $|U_{\mu h} U^*_{e h}|$ of a
heavy neutrino $\nu_h$ with $\nu_{e,\mu}$ flavors in the mass region 
$m_h \gg 1$ GeV as
derived from the sensitivity of 
present and near future $(\mu^-,e^-)$ conversion experiments.}
\label{table.2}
\begin{tabular}{lccccccc}
 &  \multicolumn{3}{c}{ Present limits} & &
\multicolumn{3}{c}{~~~~~~~~~~ Expected limits} \\ \cline{2-4} \cline{6-8} 
 Nucleus &   
           $R^{exp.}_{\mu e}$ & Ref. & $|U_{\mu h} U^*_{e h}|$ & & 
           $R^{exp.}_{\mu e}$ & Ref. & $|U_{\mu h} U^*_{e h}|$ \\ \hline 
 $^{27}Al$&  &  & & &   
 $<~ 5\times 10^{-17}$ & ~\cite{meco} & $<~ 1.2\times 10^{-3}$ \\ 
 $^{48}Ti$  & $<~ 6.1 \times 10^{-13}$ & \cite{sind} & $<~9.8\times 10^{-2}$ & & 
 $<~ 1\times 10^{-18}$ & ~\cite{prime} & $<~ 1.3\times10^{-4}$ \\ 
 $^{197}Au$  & $<~ 2.0 \times 10^{-11}$ & \cite{sind} & $<~ 0.21$ & & 
 $<~ 6\times 10^{-13}$ & ~\cite{sind} & $<~ 3.5\times 10^{-2}$ \\
\end{tabular}
\end{table}


\end{document}